\documentclass[aps,twocolumn,prl,showpacs,floatfix,superscriptaddress]{revtex4}

\usepackage{amsmath,fixmath}
\usepackage{graphicx}
\usepackage{acronym}
\usepackage{amssymb}
\usepackage{bm}
\usepackage{draftcopy}
\begin{document}

\title{First order transition in a three dimensional disordered system}

\author{L.~A.~Fern\'andez} \affiliation{Departamento
  de F\'\i{}sica Te\'orica I, Universidad
  Complutense, 28040 Madrid, Spain.}  
  \affiliation{Instituto de Biocomputaci\'on y
  F\'{\i}sica de Sistemas Complejos (BIFI), Spain.}

\author{A.~Gordillo-Guerrero} \affiliation{Departamento de
  F\'{\i}sica, Universidad de Extremadura, 06071 Badajoz, Spain.}
\affiliation{Instituto de Biocomputaci\'on y
  F\'{\i}sica de Sistemas Complejos (BIFI), Spain.}  

\author{V.~Mart\'\i{}n-Mayor} \affiliation{Departamento de F\'\i{}sica
  Te\'orica I, Universidad Complutense, 28040 Madrid, Spain.}  
\affiliation{Instituto de Biocomputaci\'on y
  F\'{\i}sica de Sistemas Complejos (BIFI), Spain.}

\author{J.~J.~Ruiz-Lorenzo} \affiliation{Departamento de
  F\'{\i}sica, Universidad de Extremadura, 06071 Badajoz, Spain.}
\affiliation{Instituto de Biocomputaci\'on y
  F\'{\i}sica de Sistemas Complejos (BIFI), Spain.}  

\date{\today}

\begin{abstract}
  We present the first detailed numerical study in three dimensions of
  a first-order phase transition that remains first-order in the
  presence of quenched disorder (specifically, the
  ferromagnetic/paramagnetic transition of the site-diluted four
  states Potts model). A tricritical point, which lies surprisingly
  near to the pure-system limit and is studied by means of Finite-Size
  Scaling, separates the first-order and second-order parts of the
  critical line.  This investigation has been made possible by a new
  definition of the disorder average that avoids the
  diverging-variance probability distributions that plague the standard
  approach.  Entropy, rather than free energy, is the basic
  object in this approach that exploits a recently introduced
  microcanonical Monte Carlo method.
\end{abstract}
\maketitle 

The combination of phase coexistence and chemical disorder plays a
major role, for instance, in colossal magnetoresistance
oxides~\cite{MANGA}. In these situations one faces a fairly general
question: {\em which are the effects of quenched
  disorder~\cite{GIORGIO} on systems that undergo a first-order phase
  transition in the ideal limit of a pure sample?}  For $D\!=\!3$
systems, $D$ being the space dimension, we only know that disorder
somehow smoothes the transition. More is known in $D\!=\!2$, where the
effects of disorder are so strong that the slightest concentration of
impurities switches the transition from first-order to
second-order~\cite{Aize89,Card90,UNIVERSALIDAD2D}.

An useful physical picture in $D\!=\!3$ is provided by the
Cardy-Jacobsen conjecture~\cite{Card90}.  Consider a ferromagnetic
system undergoing a first order phase transition for a pure sample.
Let $T$ be the temperature while $p$ is the concentration of magnetic
sites.  A transition line, $T_\mathrm{c}(p)$ separates the
ferromagnetic and the paramagnetic phases in the $(T,p)$ plane.  In
$D\!=\!3$ a critical concentration is expected to exist,
$1>p_\mathrm{t}>0$, such that the phase transition is of the first-order
for $p>p_\mathrm{t}$ and of the second order for $p<p_\mathrm{t}$ (at
$p_\mathrm{t}$ one has a {\em tricritical point}). When $p$ approaches
$p_\mathrm{t}$ from above, the latent-heat and the surface tension
vanish while the correlation-length $\xi(T_\mathrm{c}(p))$ diverges.
The Universality Class is expected to be related with that of the Random Field
Ising Model (RFIM). However, the Cardy-Jacobsen conjecture relies on a
mapping between two still unsolved  models (in $D\!=\!3$), the (large
$Q$) disordered Potts model~\cite{WU} and the RFIM.

Numerical simulation is an important tool for theoretical
investigations in $D\!=\!3$. In this way, large portions of the
transition line $T_\mathrm{c}(p)$ were found to be second
order~\cite{Ball00,Chat01,Chat05}.  However, the study of the
tricritical point as well as that of the first-order part of the
transition line seemed hopeless.  The problem comes from the
long-tailed probability distribution functions (PDF) encountered at
$T_\mathrm{c}(p)$, when comparing the specific-heat or the magnetic
susceptibility of different samples~\cite{Chat05}.  Long tailed PDFs
follows from the standard definition of the quenched free-energy at
temperature $T$ as the average of the samples' free-energy {\em at the
  same $T$\/}~\cite{GIORGIO}, which is dominated by rare
events~\footnote{For a sample of linear size $L$, the width of the
  phase-coexistence temperature interval is $\Delta T_1\!\sim\!
  L^{-D}$~\cite{FSSFO}, while each sample's critical temperature lies
  in an interval of width $\Delta T_2\!\sim\! L^{-D/2}$ around
  $T_\mathrm{c}$~\cite{ESCALA-DES}. Hence, at fixed $T$, only a tiny fraction $\Delta
  T_1/\Delta T_2\!\sim\!  L^{-D/2}$ of the samples show phase
  coexistence.}. Furthermore, the simulation of a sample of linear
size $L$ with previous methods is intrinsically hard even for a pure
system (see~\cite{NEUSHAGER}).  In fact, previous
work~\cite{Chat01,Chat05} was limited to $L\leq 25$.

Here, we study for the first time the tricritical point separating the
first and the second order pieces of the transition line. Furthermore,
we characterize a first order transition that remains so in the
presence of quenched disorder. This has been made possible by two
alternative methods of performing the sample average that avoid
long-tailed PDFs, reproduce the correct thermodynamic limit, and
provide complementary information. Essential for this study has been
the capability of studying directly the entropy, using a recently
proposed microcanonical Monte Carlo method~\cite{Mart07} combined with
a cluster algorithm~\cite{SW}. We studied systems of size up to
$L\!=\!128$, which allowed a neat Finite-Size Scaling investigation of
the {\em elusive} tricritical point.

Specifically, we consider the site diluted $Q\!=\!4$ Potts model with
periodic boundary conditions. The spins $\sigma_i\!=\!1,\ldots,Q$ occupy
the nodes of a cubic lattice with probability $p$. We consider
nearest neighbor interaction:
\begin{equation}
{\cal H}^\mathrm{spin}=-\sum_{<i,j>}\epsilon_i \epsilon_j
\delta_{\sigma_i \sigma_j}\,.
\end{equation}
The $\epsilon_i$ are quenched occupation variables,
($\epsilon_i\!=\!0$ or 1 with probability $1-p$ and $p$
respectively)~\footnote{To reduce statistical fluctuations, we kept
  only the spins in the percolating cluster~\cite{Stauffer} that
  determine the critical behavior.  However, in the most interesting
  region ($p\approx 0.96$) this correction is extremely small.}.  The
pure system, $p\!=\!1$, undergoes a first order phase
transition~\cite{Chat05,Mart07} which is generally regarded as {\em
  very strong}.

We introduce a real-valued conjugated momentum per occupied site,
$\pi_i$~\cite{Mart07}. The total Hamiltonian is ${\cal H}\!=\!{\cal
H}^\mathrm{spin}+\sum_i \epsilon_i\pi_i^2/2$ (the internal energy
density will be $e\!=\!{\cal H}/N$~\footnote{Recall that $N$ is a
random variable so that one could use as well $\tilde e= {\cal
H}/L^D$. However, $N/L^D$ is a self-averaging quantity, which makes
the difference immaterial.}). In the canonical ensemble, $\langle
e\rangle_T\!=\!1/(2T)+ \langle{\cal H}^\mathrm{spin}/N\rangle_T$.  We
consider instead the {\em microcanonical ensemble} for the extended
model $\{\sigma_i,\pi_i\}$ at fixed $e$, and integrate out the
$\{\pi_i\}$ to obtain a Fluctuation-Dissipation formalism.  The basic
quantity is a function of $e$ and the spins, $\hat\beta\!=\!(N-2)/(Ne
-{\cal H}^\mathrm{spin})\,$.  Its microcanonical mean value
$\beta_{\{\epsilon\}}(e)\!=\!\langle\hat\beta\rangle_e$ is the
$e$-derivative of the entropy per spin, $s(e)$, for that particular
sample $\{\epsilon\}$.

\begin{figure}[h]
\includegraphics[width=0.54\columnwidth,angle=270,trim=0 0 0 0]{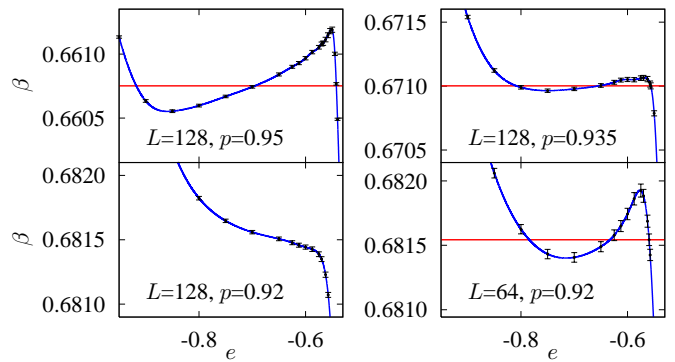}
\caption{(color online). Sample-averaged $e$-derivative of the
  entropy, $\beta(e)$, for several lattice sizes, $L$, and spins
  concentrations, $p$.  Metastability requires a non-decreasing
  $\beta(e)$. The horizontal line marks the critical (inverse)
  temperature $1/T_\mathrm{c}$, obtained through Maxwell's
  construction. At fixed $L$ the surface tension increases for growing
  $p$. Note that, for fixed dilution, a seemingly first order
  transition ($L=64$, bottom-right), may actually be of the second
  order if studied on larger lattices ($L=128$, bottom-left).  }
\label{BETAPROMEDIO}
\end{figure}

Connection with the canonical formalism is made by solving the
equation $\beta_{\{\epsilon\}}(e)-1/T\!=\!0$, that yields the internal
energy as a function of temperature.  Thermodynamic stability requires
$\beta_{\{\epsilon\}}(e)$ to be a decreasing function of $e$. Yet, at
phase coexistence and for finite $N$, it is not (see
Fig.~\ref{BETAPROMEDIO} and Ref.~\cite{Mart07}): the equation
$\beta_{\{\epsilon\}}(e)-1/T\!=\!0$ has several roots. For
$T=T_\mathrm{c}$, we name respectively $e_\mathrm{d}$ and
$e_\mathrm{o}$ the rightmost and leftmost solutions, that correspond
to the energy densities of the coexisting disordered and ordered
phases. The critical temperature is fixed by Maxwell construction: the
$e$-integral of $\beta_{\{\epsilon\}}(e) -1/T_\mathrm{c}$ from
$e_\mathrm{d}$ to $e_\mathrm{o}$ vanishes~\footnote{The Maxwell rule is
  equivalent to the standard equal-height rule for the {\em canonical}
  PDF for the energy~\cite{FSSFO}. Note that it enforces the relation,
  $s_\mathrm{d}-s_\mathrm{o}\!=\!(e_\mathrm{d}-e_\mathrm{o})/T_\mathrm{c}\,.$}.
The surface-tension, $\Sigma$, is $L^{D-1}/2$ times the integral of
the positive part of $\beta_{\{\epsilon\}}(e) -1/T_\mathrm{c}$ for
$e_\mathrm{o}<e<e_\mathrm{d}$.

\begin{figure}[h]
\includegraphics[width=0.63\columnwidth,angle=270,trim=30 10 18 0]{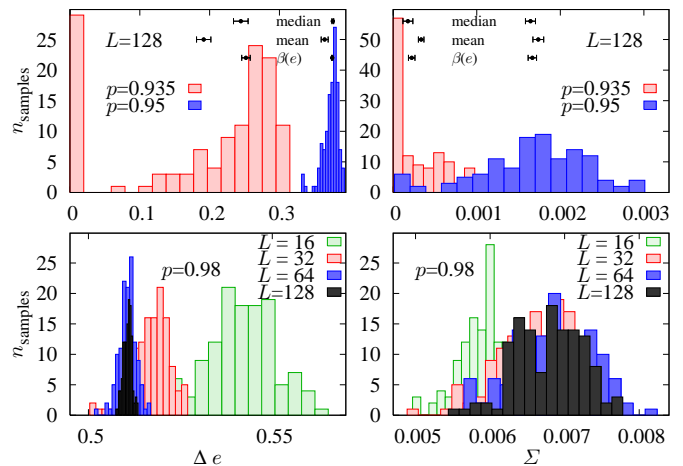}
\caption{(color online). Histograms for the sample-dependent
latent-heat $\upDelta_{\{\epsilon\}} e= e_\mathrm{d}-e_\mathrm{o}$
({\bf left}) and surface-tension ({\bf right}). In the top panels we
show results in the largest lattice, where two very close spin
concentrations behaves very differently.  The three types of drawn
horizontal lines (indicating central value and statistical error)
correspond, from top to bottom, to the median, the mean and the value
obtained from $\beta(e)$. In the lower panels we show the histograms
for $p=0.98$ and several $L$ (mind the difference in the
horizontal scales with the upper part). The latent-heat is
self-averaging while the surface tension is not.}
\label{HISTOGRAMAS}
\end{figure}

For a disordered system, one analyzes the set of functions
$\beta_{\{\epsilon\}}(e)$ corresponding to a large enough number of
samples.  There are two natural possibilities. On one hand, one can use the Maxwell
construction for each sample, extracting $T_\mathrm{c}$,
$e_\mathrm{d}$, $e_\mathrm{o}$ and $\Sigma$ and considering afterwards
their sample average or even their PDF, Fig.~\ref{HISTOGRAMAS}. The
second alternative is to compute the sample-average
$\beta(e)=\overline{\beta_{\{\epsilon\}}(e)}$, and then perform on it the
Maxwell construction (i.e. take the sample average of $s(e)$, rather
than the average of the free-energy at fixed $T$).

We have empirically found that the two sample-averaging are equivalent
in the first-order piece of the critical line. This is hardly
surprising, because the internal energy as a function of $T$ is a
self-averaging quantity for all temperatures but the critical one.
Therefore, also $e_\mathrm{d}$, $e_\mathrm{o}$ and $T_\mathrm{c}$ are
self-averaging properties in the first-order piece of the critical
line.  The first method offers more information but it is
computationally more demanding (it requires high accuracy for each
sample). The method featuring $\beta(e)$ can be used as well in the
second-order part of the critical line, nevertheless its merit in that region
are yet to be researched.

\begin{figure}[h]
\includegraphics[width=0.575\columnwidth,angle=270,trim=0 0 18 0]{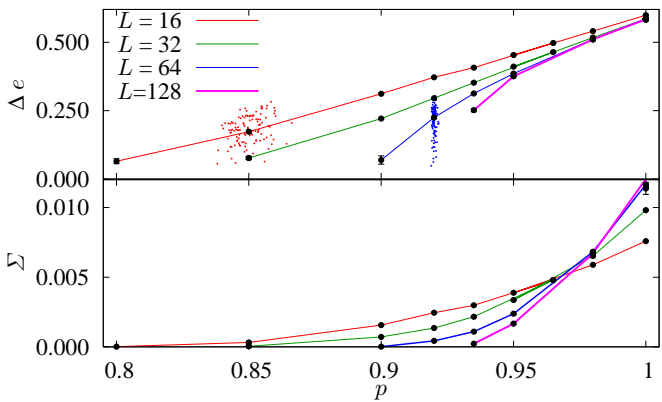}
\caption{(color online). {\bf Top:} Latent heat as obtained from
$\beta(e)$ vs. spins concentration for several lattice sizes (lines
are linear interpolations). Data for $p=1$ and $L\!=\!128$ were taken
from Ref.~\cite{Mart07}.  To illustrate the sample dispersion, we plot
as well the scatter-plot of ($N/L^D$, $\upDelta_{\{\epsilon\}}e$)
for the 128 samples at $L\!=\!16$ $p\!=\!0.85$ and $L\!=\!64$
$p\!=\!0.92$. {\bf Bottom:} as top part, for the surface tension.}
\label{LATENTE-SIGMA}
\end{figure}

\begin{figure}[h]
  \includegraphics[width=0.68\columnwidth,angle=270,trim=12 15 18
  0]{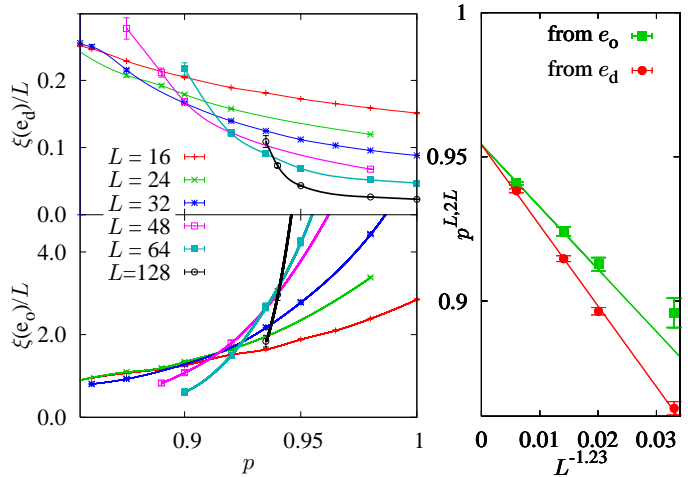}
  \caption{(color online). {\bf Left:} Correlation length in units of
    the lattice size, at phase-coexistence for the paramagnetic
    ({\bf top}) and ordered ({\bf bottom}) phases , as a
    function of spin concentration for several $L$ (lines are cubic spline
    interpolations for data at fixed $L$). {\bf Right:} Spin
    concentration where $\xi/L$ (data from left panel) coincide for
    lattices $L$ and $2L$ versus $1/L^{x}$, see
    Eqs.(\ref{SCALING},\ref{RESULTADO-GORDO}). Lines are a joint fit
    for $x$, $p_\mathrm{t}$, $A_\mathrm{d}$ and $A_\mathrm{o}$.}
\label{XIL}
\end{figure}

We have investigated the phase transition for several $p$ values in the range
$ 0.75 \leq p\leq 1$.  As a rule, we found that at fixed $p$ the
latent heat is a monotonically decreasing function of $L$,
Fig.~\ref{LATENTE-SIGMA}. For each $p$ value, we simulated $L=16$,
$32$, $64$ and $128$ (for a given $p$, we did not consider larger
lattices once the latent heat vanished). For all pairs ($L$,$p$) we
simulated 128 samples. Besides, some intermediate $L$ values were
added for the Finite Size Scaling study below (see Fig.~\ref{XIL}),
and we have raised to 512 the number of samples for ($L=16,32$,\,
$p=0.86,0.875$).

We used a Swendsen-Wang (SW) version of the microcanonical cluster
method~\cite{Mart07}. For disordered systems, SW updates properly
loosely connected regions~\cite{ISDIL} and does not require painful
parameter tunings.  For each sample, we simulated at least 20 $e$
values in the range $-1.2< e < -0.5$. The values of $e$ were decreased
sequentially, to make use of the thermalization effort at the previous
energy density.  The microcanonical cluster method, which is not
rejection-free, depends on a tunable parameter, $\kappa$.  In order to
maximize the acceptance of the SW attempt (SWA), $\kappa$ should be
chosen as close as possible to $\beta_{\{\epsilon\}}(e)$.  After every
$e$ change, we performed cycles consisting of $10^3$ Metropolis steps,
$\kappa$ refreshing, then $10^3$ SWA, and a new $\kappa$ refreshing.
The cycling was stopped, and $\kappa$ fixed, when the SWA acceptance
exceeded $60\%$.  Afterwards we performed 2---$4\times 10^5$ SWA,
taking measurements every 2 SWA.  In addition, we performed thermalization
checks that included comparisons of hot and cold starts or even mixed
configurations ({\em bands}\cite{Mart07}).

Our results for the latent-heat, $\upDelta e=
e_\mathrm{d}-e_\mathrm{o}$, and the surface tension are in
Fig.~\ref{LATENTE-SIGMA}. The apparent location of the tricritical
point (i.e. the $p$ where both $\upDelta e$ and $\Sigma$ vanish)
shifts to upper $p$ for growing $L$ rather fast. For lattice sizes
comparable with those of previous work, $L=16$, we obtain
$p_\mathrm{t}^{L=16}\approx 0.75$, at a sizeable distance from $p\!=\!1$, but 
the estimate of $p_\mathrm{t}$
increases very fast with $L$.

The PDFs for $\upDelta e$ and $\Sigma$, Fig.~\ref{HISTOGRAMAS},
display an interesting $L$ evolution.  When the $\beta(e)$ changes
behavior from non-monotonic ($L=64$, Fig.~\ref{BETAPROMEDIO},
bottom-right) to monotonic ($L=128$, Fig.~\ref{BETAPROMEDIO},
bottom-left), the two PDFs becomes enormously wide\footnote{The
estimates for $\upDelta e$ and $\Sigma$ are consistent with the {\em
median} of their (non-Gaussian) PDFs.}, see top panels in
Fig~\ref{HISTOGRAMAS}. This arises because for many $L=128$ samples,
the curve $\beta_{\{\epsilon\}}(e)$ is becoming flat, or even
monotonically decreasing (i.e. $\upDelta e\!=\!\Sigma=0$), while no such
behavior was seen for $L\!=\!64$.  Only for $p\!=\!0.98$, the width of the
PDFs for $\upDelta e$ scales as $L^{-D/2}$, as expected for a
self-averaging quantity, Fig.~\ref{HISTOGRAMAS}--bottom-left.  The
surface-tension is {\em not} self-averaging,
Fig.~\ref{HISTOGRAMAS}--bottom-right. 

From Figs.~\ref{BETAPROMEDIO},~\ref{HISTOGRAMAS}~and~\ref{LATENTE-SIGMA}
one cannot rule out that $p_\mathrm{t}=1$: a
disordered first-order transition would not exist.  Fortunately we can
solve this dilemma by considering the correlation-length, obtained
from the {\em sample-averaged} correlation function,
\begin{equation}
C(r)=L^{-D}\overline{\sum_x \epsilon_x\epsilon_{x+r}\left\langle\delta_{\sigma_x,\sigma_{x+r}}-Q^{-1}\right\rangle_e}\ ,
\end{equation}
as $\xi^2(e)=[-1+\widehat C(0,0,0)/\widehat
C(2\pi/L,0,0)]/[2\sin\pi/L]$, where $\widehat C$ is the
Fourier transform of $C(r)$ ~\cite{COOPER,AMIT}.

We take the correlation-length in units of the lattice size at
$e_\mathrm{d},e_\mathrm{o}$ as obtained from $\beta(e)$ (a jackknife
method~\cite{AMIT} takes care of the statistical correlations). For all
$p<p_\mathrm{t}$, one expects that both $\xi(e_\mathrm{d})/L$ and
$\xi(e_\mathrm{o})/L$ tend to non-vanishing and different limits for large
$L$\footnote{We have numerically checked that this is indeed the case
  for the $D\!=\!2$, $Q\!=\!4$, pure Potts model (a prototypical
  example of a second-order phase transition displaying at $T_\mathrm{c}$ a 
  double peaked
  canonical PDF for $e$).}. On the other hand, for
$p>p_\mathrm{t}$, $\xi(e_\mathrm{d})/L$ is of order $1/L$, while
$\xi(e_\mathrm{o})/L\sim L^{D/2}$. For a fixed $L$, upon increasing
$p$, the behavior goes from second-order like to first-order (see
Fig~\ref{BETAPROMEDIO}). Hence, a Finite-Size Scaling
approach~\cite{AMIT} is needed.

Consider the curves of $\xi(e_\mathrm{d})/L$ versus $p$, for different
$L$, Fig.~\ref{XIL} (left-top). There is a unique concentration,
$p^{L,2L}$, where the correlation length in units of the lattice size
coincides for lattices $L$ and $2L$. One has~\footnote{The tricritical point has no basin of
  attraction for the Renormalization Group flow in the $(T,p)$ plane.
  Although two relevant scaling fields are to be expected, the Maxwell
  construction allows us to eliminate one of them and hence we borrow
  the formula for a standard critical point.}
\begin{equation}
  p^{L,2L}\approx p_\mathrm{t} + A_\mathrm{d} L^{-x}\,,\label{SCALING}
\end{equation}
An analogous result holds for $\xi(e_\mathrm{o})/L$. Since
$A_\mathrm{d}$ and $A_\mathrm{o}$ are rather different, see
Fig.~\ref{XIL}---right, a joint fit of all data yields an accurate
estimate for the location of the tricritical point:
\begin{equation}
  p_\mathrm{t}=0.954(3),\ x= 1.23(9),\ \frac{\chi^2}{\mathrm{dof}}=\frac{4.23}{3}\,,\label{RESULTADO-GORDO}
\end{equation}
Of course, due to higher-order scaling corrections, Eq.(\ref{SCALING})
should be used only for lattices larger than some
$L^\mathrm{min}$~\cite{ON-RP2}. The fit $\chi^2$ was acceptable taking
$L^\mathrm{min}_\mathrm{o}=16$ and $L^\mathrm{min}_\mathrm{d}=12$ (for
the sake of clarity we do not display data for $L=12$ in the figures).
We thus conclude that $p=0.98$ is
definitively in the first-order part of the critical line.

We now look at $\xi/L$ at $p^{L,2L}$, Fig~\ref{XIL}.  Consider
$\xi(e_\mathrm{d})/L$ ($\xi(e_\mathrm{o})/L$) as a function of $(L,p)$, 
in the region
$p<p_\mathrm{t}$.  The salient features are: (i) for fixed $L$,
$\xi(e_\mathrm{d})/L$ is a decreasing function of $p$
($\xi(e_\mathrm{o})/L$ is increasing); (ii) for fixed $p$,
$\xi(e_\mathrm{d})/L$ has a minimum ($\xi(e_\mathrm{o})/L$ has a
maximum), at a crossover length scale, $L_\mathrm{co}(p)$, that
separates the first-order like behavior from the second order one;
(iii) at the crossing point $p^{L,2L}$ we have $L <
L_\mathrm{co}(p^{L,2L}) < 2L$; (iv) at least within the range of our
simulations, $L_\mathrm{co}(p)$ is a growing function of $p$.  A
standard scaling argument, combined with (i)---(iv), yields that
$\xi(e_\mathrm{d})/L$ at $p^{L,2L}$ is of order
$1/L_\mathrm{co}$ ($\xi(e_\mathrm{o})/L\sim
L^{D/2}_\mathrm{co}$).  If $L_\mathrm{co}(p)$ diverges at
$p_\mathrm{t}$, $\xi(e_\mathrm{d})/L$ at $p^{L,2L}$ should tend to
zero for large $L$, which is indeed consistent with our data.

In this work, we have performed for the first time a detailed study of a
disordered first-order transition in $D\!=\!3$, by site-diluting the
$Q\!=\!4$ Potts model, a system suffering a prototypically strong
first-order transition. A fairly small degree of dilution smooths the
transition to the point of becoming second order, at a tricritical
point, $p_\mathrm{t}$. A delicate Finite-Size Scaling analysis is
needed to firmly conclude that $p_\mathrm{t}<1$. We thus claim that
(quenched) disordered first-order transitions do exist in $D\!=\!3$,
although quenched disorder is astonishing effective in smoothing the
transition (we speculate that the percolative mechanism for colossal
magnetoresistance proposed in~\cite{MANGA} could be fairly common in
$D\!=\!3$). We also observe that, for a given $p<p_\mathrm{t}$, a
crossover length scale $L_\mathrm{co}(p)$ exists such that for
$L<L_\mathrm{co}(p)$ the behavior is first order like.  The
asymptotic second-order behavior appears only for
$L>L_\mathrm{co}(p)$. Our data are consistent with a divergence of
$L_\mathrm{co}(p)$ at $p_\mathrm{t}$. The successful location of the 
tricritical point has been made possible by
new definitions of the quenched average that avoids long-tailed
PDF~\cite{Chat05}. It was crucial in this approach a recently introduced
microcanonical Monte Carlo method that features the entropy density
rather than the free energy~\cite{Mart07}.

This work has been partially supported by MEC through contracts No.
FIS2004-01399, FIS2006-08533-C03, FIS2007-60977 and by CAM and BSCH.
Computer time was obtained at BIFI, UCM, UEX and, mainly, in the 
{\em Mare Nostrum.}  The authors thankfully acknowledge the computer
resources and technical expertise provided by the
Barcelona Supercomputing Center.

\end{document}